\documentstyle[galley,graphicx,epsf]{mn}
\begin{document}
\title {NGC 7419: A young open cluster with a number of very young intermediate mass pre-MS stars}
                                                                                                                      
\author[Subramaniam et al.]
        {Annapurni Subramaniam$^{1}$, Blesson Mathew$^{1}$, Bhuwan Chandra Bhatt$^{2}$, S.~Ramya$^{1}$\\
$^{1}$  Indian Institute of Astrophysics, Bangalore 560034, India\\
$^{2}$  CREST, Siddalaghatta Road, Hosakote, Bangalore}
                                                                                                                      
\maketitle
\label{firstpage}
\begin{abstract}
We present a photometric and spectroscopic study of the young open cluster NGC 7419, which is
known to host a large number of
classical Be stars for reasons not well understood. Based on CCD photometric observations of 327 stars in UBV passbands,
we estimated the cluster parameters as, reddening (E(B$-$V)) = 1.65$\pm$0.15 mag and
distance =  2900 $\pm$ 400 pc.
The turn-off age of the cluster was estimated
as 25$\pm$ 5 Myr using isochrone fits. UBV data of the stars were combined with the JHK data
from 2MASS and were used to create the near infrared (NIR)
(J$-$H) vs (H$-$K) colour-colour diagram. A large fraction of stars (42\%) 
was found to have NIR excess and their location in the diagram was used to identify them as intermediate mass
pre-MS stars. The isochrone fits to pre-MS stars in the optical colour-magnitude diagram showed that the turn-on
age of the cluster is 0.3 --3 Myr. 
This indicates that there has been a recent episode
of star formation in the vicinity of the cluster.

Slit-less spectra were used to identify 27 stars which showed H$_\alpha$ in emission 
in the field of the cluster, of which 6 are new identifications. 
All these stars  were found to
show NIR excess and located closer to the region populated by Herbig Ae/Be stars in the 
(J$-$H) vs (H$-$K) diagram. Slit spectra of 25 stars were obtained in the
region 3700\AA -- 9000\AA. The spectral features were found to be very similar to
those of Herbig Be stars. These stars were found to be more reddened than the
main sequence stars by 0.4 mag, on an average. 
Thus the emission line stars found in this cluster
are more similar to the Herbig Be type stars where the circumstellar material
is the remnant of the accretion disk. We conclude that the second episode of star
formation has lead to the formation of a large number of Herbig Be stars as well
as intermediate mass pre-MS stars in the field of NGC 7419, thus explaining the presence of
emission line stars in this cluster. This could be one of the young open clusters with
the largest number of Herbig Be stars.

\end{abstract}
\section{Introduction}
Young star clusters are formed from molecular clouds and most of the young clusters
can be found along the spiral arms which are known to trigger star formation.
A number of young star clusters are located in the Perseus spiral arm in our Galaxy, which are
naturally formed  and found due to the recent star formation in the spiral arm.
Many of the young clusters located here show the presence of a large number of classical Be stars
(h \& $\chi$ Persei, NGC 663, NGC 7419 etc.). In general, these  classical
Be stars are found to have high rotational velocity resulting in mass loss. This is
thought to be the mechanism for the presence of circumstellar material in these stars, which 
manifests as emission lines (mostly Balmer lines and some Fe lines) in their optical spectra.
The overabundance of  classical Be stars in a cluster meant that there is some mechanism at
work by which a large fraction of the early B type stars possess high rotational velocity.
One of the proposed mechanism is lower metallicity which could increase the stellar rotation.
(Maeder et al. 1999). In the case of 
NGC 146, also located in the Perseus spiral arm with 2  emission line 
stars, the cluster was found to have stars as young as 3 Myr, whereas the
turn-off age was found to be 10 -- 15 Myr. One of the  emission line stars was identified as Herbig Be star
showing signatures of accretion. Subramaniam et al. (2005) thus concluded that the Herbig Be star could have
formed from the recent star formation in NGC 146. This opened up another possibility that the clusters
in this spiral arm could have experienced continued or episodic star formation
resulting in the presence of young stars in the vicinity of the cluster. Since the molecular clouds are
abundant in spiral arms, episodic/continuous star formation in the vicinity of already formed clusters
is surely possible. Therefore, some of the Be stars found in these clusters could actually be 
similar to the Herbig Be stars, which are intermediate mass pre-MS stars.

In order to look into this possibility, it is required to estimate the duration of star formation 
in a cluster. The duration of star formation is the difference between the turn-off age (estimated from the
red giants and the MS turn-off) and the turn-on age (estimated from the pre-MS stars). 
In some clusters, the difference between the above two ages could be
a few Myr, indicating that the star formation was more or less continuous. In the case of NGC 146, 
the difference was estimated as 7 Myr, which could suggest either episodic or continuous star formation.
The age distribution of the identified pre-MS stars can also be used to differentiate between continuous
and episodic star formation.
In this paper, we study the cluster NGC 7419 in detail, to understand the
connection between the duration of star formation and the presence of  classical Be stars in this cluster.

\section{Previous studies}
NGC 7419 is a moderately populated galactic star cluster in Cepheus lying along galactic 
plane l=109.13, b= 1.14 with unusual presence of giants and super giants (Fawley \& Cohen, 1974). 
Blanco et al. (1955) identified the giants/super giants from objective-prism infrared 
spectroscopy and estimated a visual absorption of 5.0 mag and a distance of 6.6 kpc. 
van de Hulst et al. (1954) estimated a distance of 3.3 kpc,
 whereas Moffat \& Vogt (1973) estimated the distance to be 6.0 kpc.
Photometric observations of the central region of this clusters 
were done by Bhatt et al. (1993). They found a differential reddening of 1.54 to 1.88 mag with 
a mean of 1.71 mag, a cluster distance of 2.0 kpc and age about 40 Myr. Beauchamp et al. (1994) 
estimated a younger age of 14 Myr and indicated higher A$_v$ as reported by majority of authors. 
General absorption is found to be higher in this direction. Pandey \& Mahra (1987), 
Nickel \& Clare (1980) have found an absorption of 2.0-3.0 mag at 2 kpc in this direction. 
The cluster NGC7380 (l=112.76, b=0.46 and Mark 50 (l=111.36 b=-0.02 nearly in the 
direction of NGC 7419 show a range of mean A$_v$ from 1.8 to 3.4 mag (Leisawitz, 1980). 

 H$_\alpha$ emission stars in NGC 7419 were 
discovered by Gonzalez \& Gonzalez (1956) and Doldize (1959, 1975).
Further Kohoutek \& Wehmeyer (1997) updated this list with more  such stars. 
Pigulski \& Kopacki (2000) recently reported that NGC 7419 contains a 
relatively large number of  classical Be stars.  From CCD photometry in narrow band H$_\alpha$ and 
broadband R and I (cousin) filters, they identified 31  such stars. 
The fraction of  classical Be stars found in this cluster puts it along with NGC 663, which is the richest
in  classical Be stars among the open clusters in our Galaxy. 

NGC 7419 also contains a low blue-red giants ratio (Beauchamp et al., 1994). 
Caron et al. (2003) indicated a direct relation between the relative frequency of red super 
giants (RSG) stars and  classical Be stars. They found that the 
fraction of RSG increases with rotation and hence the massive stars may be rapid rotators.
Rapid rotation of early type stars is also used to explain the presence of  classical Be stars.
NGC 7419 is most likely to have solar metallicity and hence lower metallicity is unlikely
to be the reason for fast rotation.
The consensus so far, as mentioned by Caron et al. (2003), is that the stars in 
this cluster were formed from a giant molecular 
cloud with significantly higher internal motions than what is seen in the surroundings,
resulting in the formation of rapidly rotating early type stars. This suggests that the
star formation condition in this region is different from the rest of the regions
in the Galaxy, such that the the early type stars end up as rapid rotators, which is very unlikely. 
They also mention that so far, there is no direct spectroscopic estimation of the metallicity
or rotation of stars in this cluster. Hence the theory of rapid internal motions and 
different star formation process is not yet verified. Thus the problem of the high 
fraction of  classical Be stars in this cluster is still an open question. 


\section{Observations}
The photometric and the spectroscopic observations of NGC 7419 have been obtained using  HFOSC
available with the
2.0m Himalayan Chandra Telescope, located at HANLE and operated by the Indian Institute
of Astrophysics.  Details of the telescope and the instrument are available at the
institute's homepage (http://www.iia.res.in/).
The CCD used for imaging is a 2 K $\times$ 4 K CCD,
where the central 2 K $\times$ 2 K pixels were used for imaging. The pixel size is 15 $\mu$
with an image scale of 0.297 arcsec/pixel. The total area observed is
approximately 10 $\times$ 10 arcmin$^2$.
The log of observations is given in Table 1. The night of the
photometric observation was not photometric, therefore we used the previous photometric
observations of NGC 7419 (Bhatt et al. 1993) for calibrations. IRAF and DAOPHOT II routines were used
to obtain the stellar magnitudes. The error in the zero-point estimations were found
to be 0.01 mag in B and V and 0.02 in U pass band.

The cluster region was observed in the slit-less spectral
mode with grism as the dispersing element using the HFOSC
 in order to identify stars which show H$_\alpha$ in emission. 
This mode of observation using the HFOSC yields an image where
the stars are replaced by their spectra. This is similar to objective prism spectra.
The broad band R
filter (7100\AA,BW=2200\AA) and Grism 5 (5200-10300\AA, low resolution) of HFOSC CCD system was used in
combination without any slit. 
We used the Johnsons R filter in the field to restrict the spectra to the
spectral region of the R band. The image is similar to figure 2 of Subramaniam et al. (2005),
which was obtained for NGC 146. Later on, slit spectra of 25 stars identified to show H$_\alpha$ in
emission were obtained in the wavelength range, 3700\AA -- 6800\AA and 5600\AA -- 9000\AA using low resolution
grisms 7 and 8. Spectrophotometric standards such as Wolf 1346, BD 284211 and Hiltner 600
were observed on all nights. The spectra of
a known Be star, HD 58343 (B3Ve), in the blue region and a known Herbig Be star, 
HBC 705 (LKH$_\alpha$ 147, B2), in the red region were also observed and reduced in a
similar way for comparison. All the observed spectra were wavelength calibrated and corrected
for instrument sensitivity using IRAF tasks.
\renewcommand{\thetable}{1}
\begin{table*}
\centering
\caption{Log of photometric and spectroscopic observations}
\begin{tabular}{lrrr}
\hline \hline
Star/Region& Date & Filter/Grism & Exp time (sec) \\
\hline
Photometry &&&\\
\hline
N7419 & 09 July 2004&  V & 3X10, 2X60\\
      & 09 July 2004&  B & 30, 60, 2X300\\
      & 09 July 2004&  U & 120, 2X600\\
\hline
slit-less spectra&&\\
\hline
central& 30 Oct 2003& R & 5\\
       & 30 Oct 2003& R+grism5 & 120,600\\
      & 09 July 2004&  R & 5\\
      & 09 July 2004& R+grism5&900\\
north &30 Oct 2003&  R & 5\\
      & 30 Oct 2003& R+grism5&300\\
south & 30 Oct 2003&  R & 5\\
      & 30 Oct 2003& R+grism5&120,600\\
\hline
Slit spectra& & & \\
\hline
Star& Grism & Exp time (sec)& Date/Repeat date \\
\hline
N7419 B & gr7, gr8 & 900 each& 27 June 2005\\
N7419 D &      gr7 & 900 &15 July 2005 (repeat on 12 Oct) \\
        &      gr8& 1200 &15 July 2005\\
N7419 L,H &    gr7& 1200 &15 July 2005 (repeat on 08 Oct)\\
        &      gr8& 1200 &15 July 2005\\
N7419 A,I &      gr7& 1200 &15 July 2005 (repeat on 12 Oct) \\
        &      gr8& 1200 & 15 July 2005\\
N7419 C &      gr7, gr8&  900 each & 31 July 2005 \\
N7419 E &      gr7& 1200 & 31 July 2005 (repeat on 09 Oct) \\
        &      gr8& 1200 & 31 July 2005\\
N7419 J,G,3 &  gr7, gr8&  900 each & 08 Aug 2005\\ 
N7419 N,O     &  gr7&  900 & 08 Aug 2005 (repeat on 09 Oct)\\
            &  gr8&  900 & 08 Aug 2005\\
N7419 M     &  gr7, gr8&  900 each & 08 Aug 2005\\
N7419 1,2,4,5  &  gr7, gr8&  900 each & 07 Oct 2005 \\
N7419 6,P,Q,R     &  gr7,gr8&  900 each& 08 Oct 2005\\
N7419 I1    & gr7,gr8&  900 each & 09 Oct 2005 \\
N7419 F     &  gr7, gr8 & 900 each & 21 Jan 2006\\
N7419 K     &  gr7, gr8 & 900 each & 21 Jan 2006\\
\hline
\end{tabular}
\end{table*}

The field of the cluster observed photometrically is shown in Figure 1. The H$_\alpha$ emission line
stars are shown  filled circles among the normal stars (open circles). 
The pre-MS stars are identified as open circles with crosses within.
Our photometry do not have magnitudes for 5 emission
line stars and thus the locations of 22 emission line stars are shown in figure 1.
This data is used to estimate the cluster parameters as described in the 
following sections.
\section{Estimation of reddening, distance and age}
Photometric observations in the UBV pass bands are used to estimate the
reddening and distance to the cluster. There has been indications of differential
reddening across the cluster region (Bhatt et al. 1993). The (U$-$B) vs (B$-$V) colour-colour
diagram of the cluster is shown in figure 2.  The observed area as shown in figure 1 is
used to estimate the reddening and all stars within the observed region are used.
It can be seen that there is a large variation in reddening within the observed region.
The unreddened main-sequence is reddened by two values, E(B$-$V)= 1.45 and 1.85 magnitudes to
match the extreme locations of stars in the diagram.
The average reddening is found to be E(B$-$V) = 1.65 $\pm$ 0.15 mag. Assuming normal reddening law for the
interstellar reddening, the visual extinction towards the cluster is estimated as 
3.1 $\times$ 1.65 = 5.1 $\pm$ 0.4 mag. The reddening and extinction values estimated here 
are very similar to the previous estimates.

The observed colour magnitude diagram (CMD) is shown in the left panel of figure 3.
The narrow main-sequence (MS) along with the turn-off and the red giants can be clearly noticed.
In figure 2, it can be noticed that there are a few fore-ground stars which have much less reddening.
These stars are likely to cause scatter near the MS turn-off resulting in an inaccurate estimation of the
turn-off age. Therefore stars with (B$-$V) $\le$ 1.2 mag are not considered for age estimation.
The reddening and extinction corrected CMD is shown in the  right panel of figure 3.
The zero age MS (ZAMS) fit to the CMD is used to estimate the distance to the cluster. As shown in the figure,
the absolute distance modulus is estimated as 12.3 $\pm$ 0.4 mag,  which corresponds to a 
distance of 2900 $\pm$ 400 pc. The large uncertainty is due to the differential reddening across the cluster.

The MS turn-off of the cluster along with the red super giants are used to estimate the 
post-MS age using the Padova isochrones (Bertelli et al. 1994).  In our CMD, we have shown 
three red super giants and one bluer giant. This cluster is known to have 5 red super giants
and are located close to one another, and found to be members based radial velocity (Beauchamp et al. 1994).
In the right panel of figure 3,
three isochrones with ages log(age) = 7.3, 7.4 and 7.5 are shown, which correspond to
20, 25 and 32 Myr respectively. The 25 Myr isochrone is shown as bold line and the other two are
shown as dotted lines. The turn-off age of the cluster is thus found to be 
between 20 -- 30 Myr, while the 25 Myr isochrone
visually appears to fit better. Thus we estimate the age of the cluster as 25 $\pm$ 5 Myr.
The present age estimate is similar to that derived by Bhatt et al. (1993), but older than
that estimated by (Beauchamp et al. 1994).  We also estimated the age of the cluster using synthetic CMDs.
We constrained the number of red supergiants to create synthetic CMDs using PADOVA models (Bressan et al. 1993).
The constructed CMDs are similar to those shown in figure 7 of Subramaniam et al. (2005).
It was found that the best fitting CMDs were possible for ages 20 -- 22 Myr. For ages younger than 20 Myr,
the clump of red giants do not form and also, the tip of the MS is brighter than what is observed.
For ages older than 22 Myr, large number of blue supergiants are formed. Therefore, the method of
synthetic CMDs estimate the age of the cluster between 20 --22 Myr, which is similar to the age estimated
using the isochrones. Therefore, we confirm that the turn-off age of this cluster is $\sim$ 25 Myr.
Thus the cluster is not very young to be populated by intermediate mass pre-MS stars,
since these stars typically spend only a few million years in the pre-MS phase. Thus the turn-off
age does not indicate any likely hood of the presence of intermediate mass pre-MS stars in the cluster.

\section{Near IR data and pre-MS stars}
The optical data is cross correlated with the 2MASS image to  identify the common stars in 
both the data set. Of the 327 stars for which we could obtain UBV photometry, we identified
NIR counterparts for 198 stars. Thus, J,H,K magnitudes were combined with the present UBV photometry for
198 stars in the cluster region.
 The (J$-$H) versus (H$-$K) colour-colour diagram of all the stars
in the cluster region are shown in  the left panel of figure 4. 
The diagram also shows the location of MS stars, giants, and
the reddening line (Bessel \& Brett 1988). The two boxes shown are the location where classical Be stars and
Herbig Ae/Be stars are generally found (Dougherty et al. 1994 and Hernandez et al. 2005). The dashed line
indicates the location of T-Tauri stars (Meyer et al. 1997).  The reddening corrected figure is shown in 
the right panel. The value of E(J$-$H) used for reddening correction is 0.45 mag and E(H$-$K) is 0.23 mag.
The black triangles indicate optically identified stars and the 
triangles with open circles indicate
stars with H$_\alpha$ in emission. The foreground stars are not shown in this figure.

Stars (denoted by triangles) located to the right of the reddening line have NIR excess.
It can be seen that the whole cluster sequence is highly reddened with the presence of a large number of stars
with NIR excess. 83 stars were found to show NIR excess among the 198 stars. 
The emission line stars also show NIR excess, except star K (see table 2).
Due to heavy reddening, the emission line stars are 
not found near the Be star box, but are distributed close to the Herbig Ae/Be location and a few of them within the 
Herbig Ae/Be box.  The panel on the right side shows the de-reddened sequence. Some of the stars
show negative colour due to over correction of the reddening, due to variable reddening in the region.
It can be seen that the some emission line stars are now located in the Be box, though a large
fraction still lies outside. The stars infact are located in a region which connect the loci of
classical and Herbig Be stars, with a couple of emission line stars located within the Herbig Ae/Be box.
It is necessary to find the evolutionary status of these stars in order to
understand their peculiar location in the diagram. 
It can also be seen that the stars with NIR-excess are located below the dashed line (just above the
location of Herbig Ae/Be stars), which denote the location of T-Tauri stars. 
Thus, the optically identified stars with NIR excess are most likely to be 
intermediate mass pre-MS stars. If we extend this arguement to the emission line stars, then
it is likely that these stars also could belong to the same category. We find that the fraction
of pre-MS stars is 42\%, which is almost half of the stars in this cluster. 
These pre-MS stars are very likely to
be very young and their age is estimated in the following section.

\section{Pre-MS age estimation}
In general, the right side of the MS at the fainter
end of the cluster CMD is populated with field giants. But this is the same region which is occupied by the
pre-MS stars, if they are present in the cluster. In the previous section, we
identified some stars to have NIR excess which have optical counterparts.
If these are pre-MS stars, then they should be located to the
right of the MS in the optical CMD. In other words, the presence of NIR excess, and their location in the
NIR colour-colour diagram can be used to identify the pre-MS stars from the field stars in the 
right side of the CMD. The cluster CMD is shown in figure 5, where only stars with NIR excess 
are shown and the emission line stars are shown as dots with circles 
around them. It can be seen that most of these stars are located to the right side of the
MS. This confirms that the stars with NIR excess are pre-MS stars.
Also, a large fraction (42\%) of stars are in the pre-MS phase. 
In order to estimate the age of these pre-MS stars,
pre-MS isochrones from Siess et al. (2000) are plotted on the figure.
The isochrones shown are for ages, 0.2, 0.3, 0.4, 0.5, 1.0, 1.5, 3.0, 5.0
and 10 Myr in dashed lines.. The post-MS isochrone of age 25 Myr and ZAMS are also shown as 
continuous lines.  Emission line stars, which have V$_0$ $\sim$ 10.0 mag and fainter are
well matched by the isochrones of ages 0.3 - 1.0 Myr.
It can be seen that, the clump of emission line stars located at (B$-$V)$_0$ $\sim$ 0.0 mag and 
V$_0$ $\sim$ 11.0 mag is located just brighter than the 0.3 Myr pre-MS isochrone  and the
isochrones do not reach this location. One possible reason is that these stars are more massive 
than 7.0 M$_\odot$, which is the upper mass limit of the
model used. These stars could also be younger than 0.3 Myr. 
Therefore, all the emission line stars
are found to have ages less than 1.0 Myr. A number of stars with NIR, which did not show any H$_\alpha$
emission are found to have ages between 0.5 and 3.0 Myr. Only two stars are found to be older than 3.0 Myr.
The limiting magnitude
of the CMD is V$_0$ $\sim$ 15.0 mag and only two stars are found in the magnitude range of 13.0 - 15.0 mag.
Thus it is very likely that there are very few pre-MS stars with ages older than 3.0 Myr.
 
Thus the 42\% of the stars in the cluster is likely to have
formed in the last 3 Myr or less than that.

Since we have estimated the turn-off and the turn-on age, we shall now discuss the
{\it star formation history} of the cluster.
The values derived for the age of the cluster from the turn-off and the turn-on
are quite different, since
the post-MS age of the cluster is found to be much higher than
the pre-MS age (25 and 3 Myr). It is very unlikely that the pre-MS and the turn-off stars
belong to the same population, as the star formation in the cluster is unlikely to
have continued  for more than 20 Myr. It is more likely to come 
from two different bursts of star forming events.
The presence of the pre-MS stars can be understood if the
cluster has experienced a recent event of star formation in its vicinity.
This star formation is likely to have happened in the last 3 Myr with most of
the massive stars forming in the last 0.3 Myr.
Thus we propose that NGC 7419 cluster region experienced two episodes of star formation
one around 25 Myr and the second between 0.3 -- 3 Myr.
The pre-MS isochrone fits
clearly point out that at least 12 stars with H$_\alpha$ emission have age as young as 0.3 Myr.
The brighter emission line stars are also likely to belong to this group.
In the optical CMD all the emission line stars are more reddened than the MS.
Also, in the NIR colour-colour diagram, they are not located in the location for Be stars, 
but in a region with larger 
reddening and NIR excess and thus closer to the location of Herbig Ae/Be stars. Combining
all the above, it is very clear that the so called classical Be stars in this cluster are more likely
to be very young pre-MS stars in the B-spectral type. Thus these stars are similar to the
Herbig Be stars in their evolutionary state and not to the fast rotating classical Be stars.
In the following section we present the spectra of the emission line stars and compare them with
the spectra of classical and Herbig Be stars.

\section{Emission line stars}
Pigulski \& Kopacki (2000) found 31 Be stars in the central region of the cluster using
H$_\alpha$ photometric observations. In order to confirm their identification, we obtained
slit-less spectra of the central as well as the northern and southern regions.
The spectra of stars brighter than
V=16.0 mag could be identified by this technique. This technique helps to identify stars with
medium to strong H$_\alpha$ emission. When there is a fill up of the absorption line due to
emission, this technique fails to identify such stars. Also, near the center of the cluster, due
to crowding effects, spectrum from a fainter star gets submerged within that of a brighter star.
Thus, this technique picks up genuine and bona-fide H$_\alpha$ emission stars and could result in incomplete
detection of the emission line stars due to the above mentioned problems. Slit-less spectra
were obtained on two- three epochs in order to identify  emission stars which could have been
missed out due to variable H$_\alpha$ emission. After analysing these images, we could detect
21 emission line stars in the same region as studied by Pigulski \& Kopacki (2000), against their
detection of 31 stars. Outside the central region, we detected 6 more stars thus increasing the
total number of detections to 27. We do not detect emission in 10 stars which are previously
found to show emission. Of these, 4 stars were found to be too faint to detect their spectrum,
4 stars were found to show continuum spectrum without any lines and two stars were found to show
H$_\alpha$ in absorption. Photometric information of the 27 emission stars are tabulated in table 2.
R and (R$-$I) colour are taken from Pigulski \& Kopacki (2000). Reddening to 
individual stars is estimated from their spectra. 

The location of the emission line stars in the optical CMD indicate that they are brighter than
M$_V$ = $-$ 2.0 mag, thus these stars are early B type stars. We have obtained UBV photometry of 22
emission line stars, which are shown in figure 1.
We do not have the UBV photometric data for stars F, I1 and P, due to high photometric 
errors and stars S and T are out of the field of photometric observation. 2MASS J,H,K data is either
not available or have high errors for stars 1 - 6.
In order to understand the nature of these 
emission line stars, we obtained slit spectra in the range 3800 \AA -- 9000 \AA\ for 25 emission line stars.
The blue part of the spectrum is shown in figures 6, 7 and 8. For comparison, we have also shown the
spectrum of a Herbig Be star HBC 705 (spectral type B3, shown in red), which is taken from Hernandez et al. (2004). 
It can be seen that the spectral features and the slope of the continuum are very similar.
Also shown is the spectrum of HD 58343, a known Be star of spectral type B3 and having 
V$_{rot} < 40$ kms$^{-1}$ (Slettebak et al. 1992).  This is shown in blue colour in the figures.
 
The striking feature of the spectra of emission line stars is its slope. Just the slope of these
B type stars suggest that they belong to the late type, in sharp contrast to the spectrum of the classical
Be star, HD 58343.  A comparison with the spectrum of HBC 705
reveals that it could be of a B type star with very high extinction. Thus the spectra in the blue region, of 
the emission line stars indicate that all have high extinction.

\renewcommand{\thetable}{2}
\begin{table*}
\begin{center}
\caption{Photometric data of the identified emission line stars are tabulated. The first column
represents the identification used here, the second column represents their number in Beauchamp et al. (1994).
Columns 3 -- 5 is our data, 6 --7 are taken from Pigulski \& Kopacki (2000). 8 - 10 are from 2MASS catalogue
and the reddening in the last column is derived from their blue spectra. The error in the estimated reddening is
0.1 - 0.2 mag.}
\end{center}
\begin{tabular}{lrrrrrrrrrrrr}
\hline
our no.&  BMD& X & Y & V& (B$-$V)& (U$-$B)&        Rc  &  (R-I)c&  J &     H&      K&   E(B$-$V) \\
       &   & pixel& pixel&mag& mag&              mag&   mag&   mag&   mag&    mag&   mag& mag \\
\hline
A    &  1076 & 1438.18 & 1045.17 &14.34 &1.64 &0.55&          13.21 &  1.15 & 10.41 &  9.81  & 9.36& 2.2\\
B    &   831 & 1203.74 & 989.98  &14.77 &1.53 &0.51&          13.78 &  1.03 & 11.41 & 10.80 &10.26 & 1.9 \\
C    &   781 & 1167.58 & 957.21  &14.06 &1.43 &0.52&          12.89 &  1.09 &  9.90 &  9.27 & 8.79 & 2.0 \\
D    &   728 & 1120.37 & 858.29  &15.56 &1.50 &0.57&          14.78 &  1.01 & 12.22 & 11.66 &11.19 & 2.1 \\
E    &   745 & 1132.46 & 1245.57 &16.56 &1.70 &0.69&          15.39 &  1.18 & 12.44 & 11.75 &11.25 & 2.0 \\ 
F    &   741 &         &         &      &     &    &          16.08 &  1.21 & 13.09 & 12.37 &11.89 & 2.3 \\
G    &   620 & 1040.42 & 787.60  &14.25 &1.54 &0.54&          13.12 &  1.14 & 10.36 &  9.70 & 9.21 & 1.9\\
H    &   582 & 1017.98 & 1044.08 &15.67 &1.45 &0.63&          14.63 &  1.02 & 11.61 & 11.04 &10.64 & 2.1\\
I    &   585 & 1017.98 & 1044.08 &14.24 &1.51 &0.54&          13.44 &  1.02 & 11.03 & 10.53 &10.16 & 2.1 \\
I1   &   OUT &         &         &      &     &    &                &       &       &       &      & 2.4 \\
J    &   417 & 913.54  & 792.92  &13.93 &1.62 &0.61&          13.14 &  1.06 & 10.46 &  9.94 & 9.64 & 2.0 \\
K    &   621 & 1039.66 & 989.00  &14.79 &1.51 &0.61&          13.93 &  1.12 & 11.33 & 10.82 &10.59 & 2.3 \\
L    &   504 & 975.81  & 971.92  &15.09 &1.54 &0.55&          14.08 &  1.11 & 11.43 & 10.85 &10.36 & 2.2 \\
M    &   389 & 889.31  & 1031.15 &13.72 &1.60 &0.60&          12.64 &  1.14 &  9.95 &  9.32 & 8.84 & 2.1 \\
N    &   427 & 923.62  & 1200.29 &15.37 &1.74 &0.70&          14.03 &  1.23 & 10.98 & 10.29 & 9.77 & 2.2 \\
O    &   232 &         &         &      &     &    &          15.51 &  1.33 & 12.12 & 11.35 &10.84 & 2.2 \\
P    &   OUT & 878.74  & 1500.79 &15.36 &1.63 &0.73&                &       & 11.50 &10.89 &10.47 & 2.1 \\
Q    &   OUT & 903.56  & 483.38  &15.97 &1.64 &0.62&                &       & 12.18 &11.55 &11.09 & 2.2 \\
R    &   OUT & 959.03  & 431.47  &16.09 &1.48 &0.65&                &       & 12.49 &11.89 &11.49 & 2.2\\
S    &   OUT &         &         &      &     &    &                &       & 12.07 &11.31 &10.85 & -\\
T    &   OUT &         &         &      &     &    &                &       & 12.93 &12.21 &11.69 & - \\
1    &   884 & 1257.71 & 936.35  &15.73 &1.53 &0.58&          14.64 &  1.07&        &      &      & 2.1\\
2    &   815 & 1188.64 & 945.16  &15.35 &1.38 &0.55&          14.58 &  1.02&        &      &      & 2.2 \\
3    &   692 & 1093.24 & 783.30  &14.19 &1.62 &0.57&          13.05 &  1.15&        &      &      & 1.8\\
4    &   702 & 1100.23 & 748.24  &14.17 &1.62 &0.59&          13.15 &  1.13&        &      &      & 2.1 \\
5    &   239 & 722.58  & 682.02  &15.40 &1.87 &0.86&          13.90 &  1.31&        &      &      & 2.3 \\
6    &   795 & 1174.97 & 1173.92 &16.26 &1.59 &0.72&          15.02 &  1.17&        &      &      & 2.1\\
\hline
\end{tabular}
\end{table*}

The prominent lines in the spectrum are indicated. The presence of 4026, 
4121 and 4388 \AA He I absorption lines
in the spectra indicate that these stars are indeed early type B stars. 
This supports the photometric spectral type derived based on the estimated
distance. 
There are a number of features present in the spectra, which are similar to that in the
spectrum of the Herbig Be star. This striking similarity of the spectra with that of
a Herbig Be spectra than that of a Be star spectrum supports the earlier result that
these stars are very young pre-MS stars. The width of the H$_\gamma$ line is similar to that
of the Be star, indicating that the rotation velocity is similar. In some stars, this
line is found to show structures indicating that there could be some fill up in the
absorption. The H$_\beta$ line is found to be in emission in most 
of the stars. 
This part of the spectrum is
used to estimate the reddening towards each star, by de-reddening the spectra. The comparison
spectrum used is a B3 star spectrum from the spectral library. These reddening estimates are
shown in table 2. The error in the estimate is 0.1 mag for brighter stars and 0.2 for fainter stars
where the spectra are noisy. On an average the emission line stars have reddening of 2.1 mag, which is
$\sim$ 0.4 mag more than that estimated for the MS stars. This indicates that the emission line stars
have circumstellar material with a visual extinction of $\sim$ 1.2 magnitude. 
The spectra of stars F, I1, K, O and P are not shown since the spectra are very noisy.

 Spectra of some stars in the region 5800 - 6800 \AA~
are shown in figures 9. Strong H$_\alpha$ emission can be noticed in all the spectra.
A number of Fe II lines can be seen in emission and these are indicated in the figure.
The  H$_\alpha$ emission line does not show prominent wings and the structure is very similar to that
seen in HBC 705 (spectrum shown in red colour). In this part of the spectrum also, the spectra shows remarkable
similarity to that of the Herbig Be star. The He I (6678 \AA) is seen in absorption, but shows
structures in the profile in some stars, indicating that there is emission within the absorption.
This feature is also seen in the spectrum of HBC 705. 

The red spectra of these stars indicate the presence of a few more Fe emission lines. O I (7773 \AA)
is found to be in emission in most of the stars. The red spectrum in the region of 
Ca II triplet and Paschen lines
are shown in figures 10 and 11. The spectrum of HBC 705, obtained using HCT is over plotted in red.
Various spectral lines are indicated in the figure.
In most of the stars, O I (8446 \AA) and the Paschen lines are in emission. 
The presence of Ca II triplet cannot be
identified since they get merged with the emission lines of the Paschen series. In the 
case of star I, the spectrum shows faint Paschen emission lines, whereas the emission
lines near the Ca II triplet is stronger which may indicate a significant contribution 
from Ca II triplet. These emission profiles are found to show structures.
Also the Paschen emission lines in 12 stars (C,G,E,B,I,N,L,3,4,2,1 and Q) are found to show structures, 
like in the spectrum of HBC 705. In stars, 5,M,J,O,A and D, single peaked strong emission could be seen.
The emission profiles indicate the presence of disk around the star. Multiple emission peaks
indicate velocity structure/gradient in the disk. Since these emission line stars
belong to the early B type, the presence of low ionisation lines indicate that either these stars have
a large amount of circumstellar material or that the disk is quite extended. Since the emission profiles are
quite sharp, large velocity gradient in the disk unlikely to be present. Estimation of large 
circumstellar reddening (0.2 -- 0.6 mag)
supports the suggestion that these stars have a large amount of
material around them. This again indicates that these stars are similar to the Herbig Be type stars
than the classical Be stars. The stars F and I1 show no absorption or emission features, whereas
H, K, P, R and 6 show a few absorption features in this region of the spectrum.
Caron et al. (2003) presented high resolution spectra of M (389), 
J(417) and G(620) in the region 8300 -- 8900 \AA. They also find strong emission in Paschen lines
in the spectra of M and G, with a lot of structure in the emission lines of G. We see only a marginal evidence
of structure in our spectrum of star G, but it is seen very clearly in the high resolution spectrum
of Caron et al. (2003). The star J (417) displays weak emission, similar to a shell, in both the
low and high resolution spectra. Thus the features identified here are in complete agreement with
high resolution spectra in the red region presented in Caron et al. (2003).

From the spectra presented here, it is clear that the emission stars in this cluster have features 
very similar
to those found in a Herbig Be star than in a classical Be star. Also, the spectra indicate the presence
of circumstellar disk and high extinction. No indication of high rotational velocity, either in 
the stellar absorption
lines or in the circumstellar emission lines is seen. This result in combination with the presence
of NIR excess, high circumstellar reddening and a very young pre-MS age, strongly suggests that
these stars are Herbig Be type stars. Thus the circumstellar disk
is the remnant of the accretion disk of the pre-MS phase and not formed due to fast rotation. 

\section{Components of reddening towards the cluster}
The reddening estimated towards the normal and the emission line stars can arise due to a number of
components located in the line of sight. In the case of this cluster, we were able to de-convolve the
various components. The cluster can be reddened due to (a) {\it interstellar} material located between the
cluster and us, (b) the {\it circumcluster} material which is left behind after the star formation event
and (c) the {\it circumstellar} material around the pre-MS stars.
We estimate the three components mentioned above. The clusters NGC 7510 (l=110.90$^o$ and b=0.06$^o$) and King 10
(l= 108.48$^o$ and b = $-$0.40$^o$) are young clusters (30 and 25 Myr respectively) located in 
same direction of NGC 7419.  They also have similar
distance moduli (12.5 and 12.6), but located marginally farther away. 
The estimated value of reddening towards these clusters is E(B$-$V) = 1.0 mag.
Thus, the reddening to similarly aged clusters located at similar distance in the same direction
is less by 0.65 mag. NGC 7226 is another cluster located in the same direction but slightly older, 300 Myr.
This cluster is located at 2 kpc and the estimated reddening towards this cluster is 0.5 mag. NGC 7245,
located at 2 Kpc in the same direction is also found to have a reddening of 0.5 mag. Thus it can be seen that
in this direction, the reddening increased between 2 and 3 kpc by 0.5 mag, whereas up to 2 kpc it is only
0.5 mag. Thus we estimate that the reddening due to the component (a) is 1.0 $\pm$ 0.15 mag. The estimated 
reddening to
NGC 7419 is 1.65 mag, which is higher than that found for this distance. 
The fact that the cluster stars are more reddened indicates that some material is present in the immediate 
vicinity of
the cluster. Thus the reddening due to the circumcluster material is 0.65 $\pm$ 0.15 mag. Thus we find that there is
substantial reddening just around the cluster and this supports the result that the cluster region
experienced a recent episode of star formation. Thus the estimate for the second component of
reddening (b) is 0.65 $\pm$ 0.15 mag. Now the emission line stars are found to have NIR excess and higher reddening.
We estimated the third component (c) for these stars using their spectra. The average value of the
circumstellar reddening for the emission line stars is found to be 0.4 $\pm$ 0.1 mag, as estimated
in the previous section. To summarise, the reddening
decomposition shows that there is significant {\it circumcluster} material, supporting the idea of
a recent star formation
event in the cluster region. The emission line stars are thus more likely to be formed as a result
of this star formation event and are early type pre-MS stars, with substantial circumstellar reddening.

\section{XMM observations}
Early B type stars can show H$_\alpha$ in emission, if they are in  
accreting binary systems. There was a proposal by Christian Motch to detect X-ray emission from the 
emission line stars in this cluster and NGC 663.
The proposal was to search for Be + white dwarf binaries in a sample of Be-rich young
open clusters rich in Be stars. 
These two clusters are found to have a large fraction of Be stars among the B stars and it was suspected that
many may be accreting binary systems. XMM has the capability to efficiently
detect white dwarfs accreting from the circumstellar envelopes of Be stars.
This cluster was observed by XMM on 2004-02-02 and the data became public on 2005-04-16.
There are 63 sources detected in the direction of NGC 7419.
By comparing with the optical data, we find that none of these detections coincide with the Be stars. 
A few optically faint sources are found to be detected in X-ray and these may be low mass pre-MS stars like
T-Tauri stars. A number of sources have also been detected outside the radius of this cluster. Thus we conclude
that the X-ray flux from these stars are below the detection limit of XMM observations.
Since these observations were expected to detect X-ray binaries at this distance, the non-detection
shows that none of these Be stars are accreting binaries.

\section{Results and Discussion}
The optical and NIR photometry of stars in the region of the young cluster NGC 7419 is used to
estimate the cluster parameters. The estimated values of reddening and distance towards the cluster, and
the turn-off age are found to be similar to the previous estimates. By combining the optical and NIR
photometry, we identified the presence of a large number of stars (83 stars, 42\%) with NIR excess. These
stars were found to be intermediate mass pre-MS stars in accordance with their location in the
NIR colour-colour diagram and optical CMD. The age of these pre-MS stars is found to be between 
 0.3 -- 3 Myr,
whereas the age of the cluster as obtained from the turn-off is found to be 25 $\pm$ 5 Myr.
These two ages are possible for the same cluster, if the cluster region experienced two episodes of
star formation, one at 25 Myr and the other at about 3 Myr. We find that 42\% of stars in this
region could belong to the recent star formation episode. The slit-less spectra of the cluster region
revealed the presence of 27 emission line stars, where 6 are new detections.  All the emission line
stars (except one) have NIR excess and are found to have ages less the 1 Myr, with some as young as 0.3 Myr.
Analysis of the slit spectra
of 25 stars showed that the spectra closely resemble a Herbig Be star spectrum than a classical Be star spectrum.
The spectra show evidence for the presence of circumstellar disk and
a large amount of circumstellar material. The reddening estimated from the
spectra showed that the emission line stars are more reddened by 0.4 mag than the normal cluster stars. Thus we
conclude that the emission line stars found in this cluster are more likely to be Herbig Be type stars,
which are the early type pre-MS stars and not classical Be stars. Also, XMM observations showed that the
Be stars in this cluster are not likely to be accreting binary systems. The cluster is also found to have
circumcluster reddening of 0.65 mag supporting the evidence for a recent star formation event in the
cluster region. By definition, Herbig Be stars are those which have nebulosity around them. In this case,
we could not detect any nebulosity, may be because the cluster is located at large distance.

As shown in figure 1, the pre-MS stars are found to be located uniformly across the cluster.
On the other hand, the Be stars are found to be located closer to the cluster center. This might
indicate that the more massive stars are formed preferably towards the cluster center. 
We are unable to quantify this
statement since stars formed in two episodes cannot be separated. In this cluster we find that the second
episode of star formation has occurred between 0.3 - 3 Myr. 
Since the upper age limit of the emission line stars is 1 Myr, we can conclude that
the accretion disk for the early B type stars can survive at least up to 1 Myr.
This cluster is most probably the only known open cluster
with the largest number of Herbig Be stars. This cluster thus gives a rare opportunity to understand the
accretion disk around intermediate mass pre-MS stars as a function of their mass.
 It is likely that there are high mass ($\ge$ 7 M$_\odot$) stars in the pre-MS phase, which are
also as young as 0.3 Myr. Therefore, this cluster is very ideal to test the pre-MS evolutionary models
of stars in this mass range.

The presence of a large number of  classical Be stars among the early type B 
stars in this cluster has been a puzzle.
In this paper we argue that the emission line stars present in this cluster are not classical Be stars, but
Herbig Be stars. In the earlier sections, we found that the cluster requires some special property like
lower metallicity or  chaotic cloud environment to explain a large fraction of Be stars. In the present
scenario, no such special property is required for the cluster or the environment. 
Since the cluster is located in the
Perseus spiral arm, multiple star formation episodes are possible in the vicinity of the cluster.
The implication of the present result is that
in young open clusters, the  emission line stars may not have been identified properly. 
There could be genuine  classical Be stars, but
there is also the possibility that they are not fast rotators, but those with the remnant of the accretion disk.
In this cluster, we find that the  emission line stars populate the region between the location for 
classical and Herbig
Be stars in the NIR colour-colour diagram. The spread in their distribution in the figure is most 
likely due to the difference in the 
amount of circumstellar material around them. The distribution of their location in the diagram points to the
possibility that with the evaporation of the disk, these stars can gradually come down along the reddening
vector, to populate the region of classical Be stars. If this is the case, then can some of the stars
located in the classical Be region be evolved Herbig Be stars? In short, can the Herbig Be stars evolve to become
a  classical Be star without fast rotation?  If possible, then what is the time-scale required? 
We plan to study more open clusters with classical Be stars in order to address this question. 

The present study has demonstrated that it is very important to combine the optical and NIR data for young
open clusters to understand the properties of the cluster. The NIR data in combination with the
optical data can be effectively used as a tool to differentiate pre-MS stars from the field stars.
It is also very essential to derive not only the turn-off age, but also the turn-on age using the
pre-MS stars. These two ages are required to identify the presence of any continued/episodic
star formation in the vicinity of the cluster. We also find that continued/episodic star formation
around clusters located on spiral arms are not uncommon. Another point to look into is the effect of
episodic/continued star formation from the context of initial mass function. When the initial mass
function of a cluster is estimated, one generally assumes that the cluster is formed in one single burst
of star formation and that all the stars have almost the same age. We find that this assumption need not be
correct (eg. NGC 146) and in some cases, like NGC 7419, it can be completely wrong. Continued or
episodic star formation in the cluster region can lead to the presence of pre-MS stars and thereby
to an inaccurate estimation of the mass function slope.
Therefore, understanding the {\it star formation history} of young clusters is necessary before one attempts to
to estimate various cluster parameters.


{}

%
%
\begin{figure*}
\epsfxsize=18truecm
\epsffile{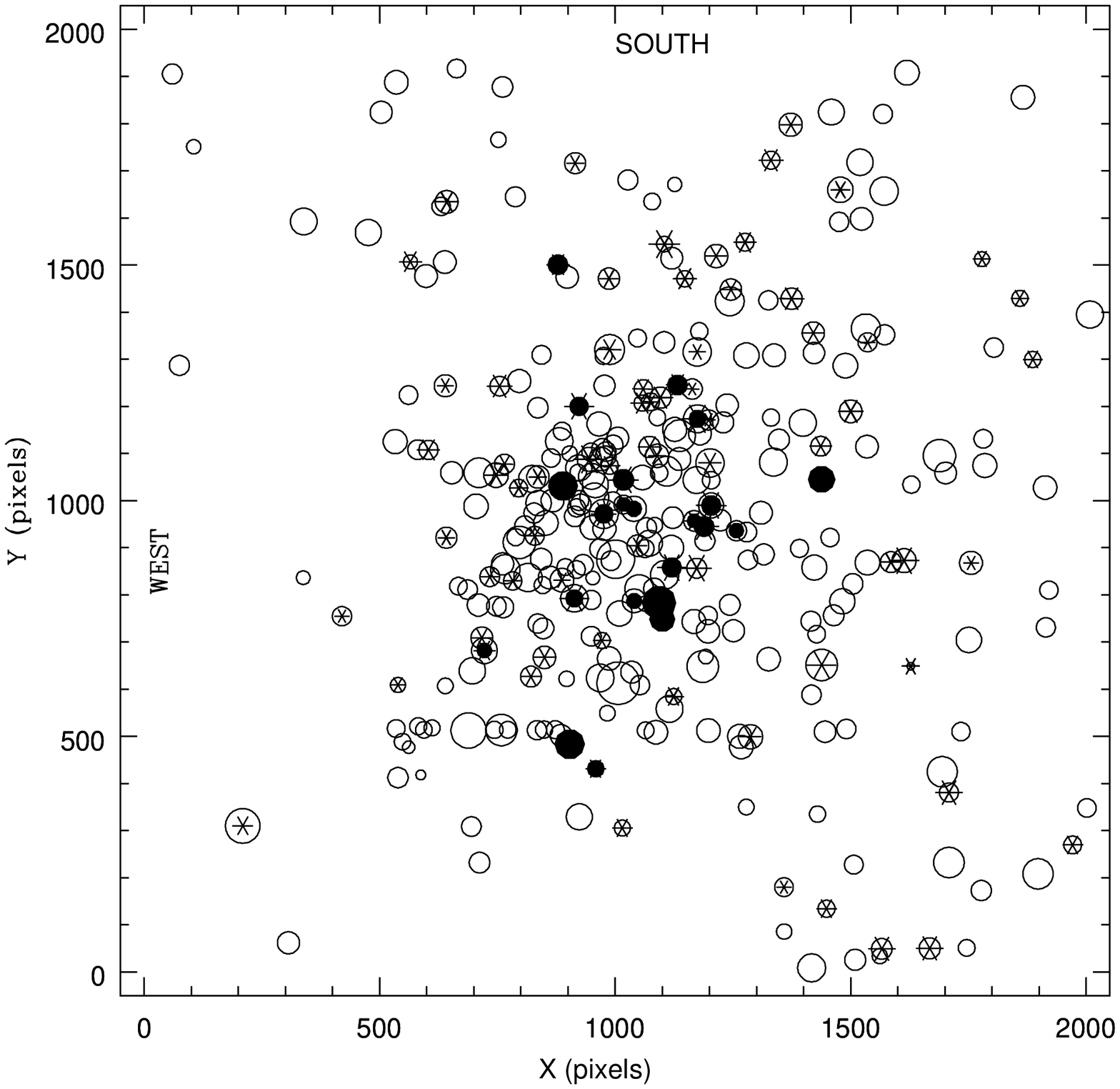}
\caption{The observed region of the cluster NGC 7419. North is down and East is to the right. 
Crosses within open circles are stars with NIR excess (pre-MS stars).}
\end{figure*}

%
%
\begin{figure*}
\epsfxsize=18truecm
\epsffile{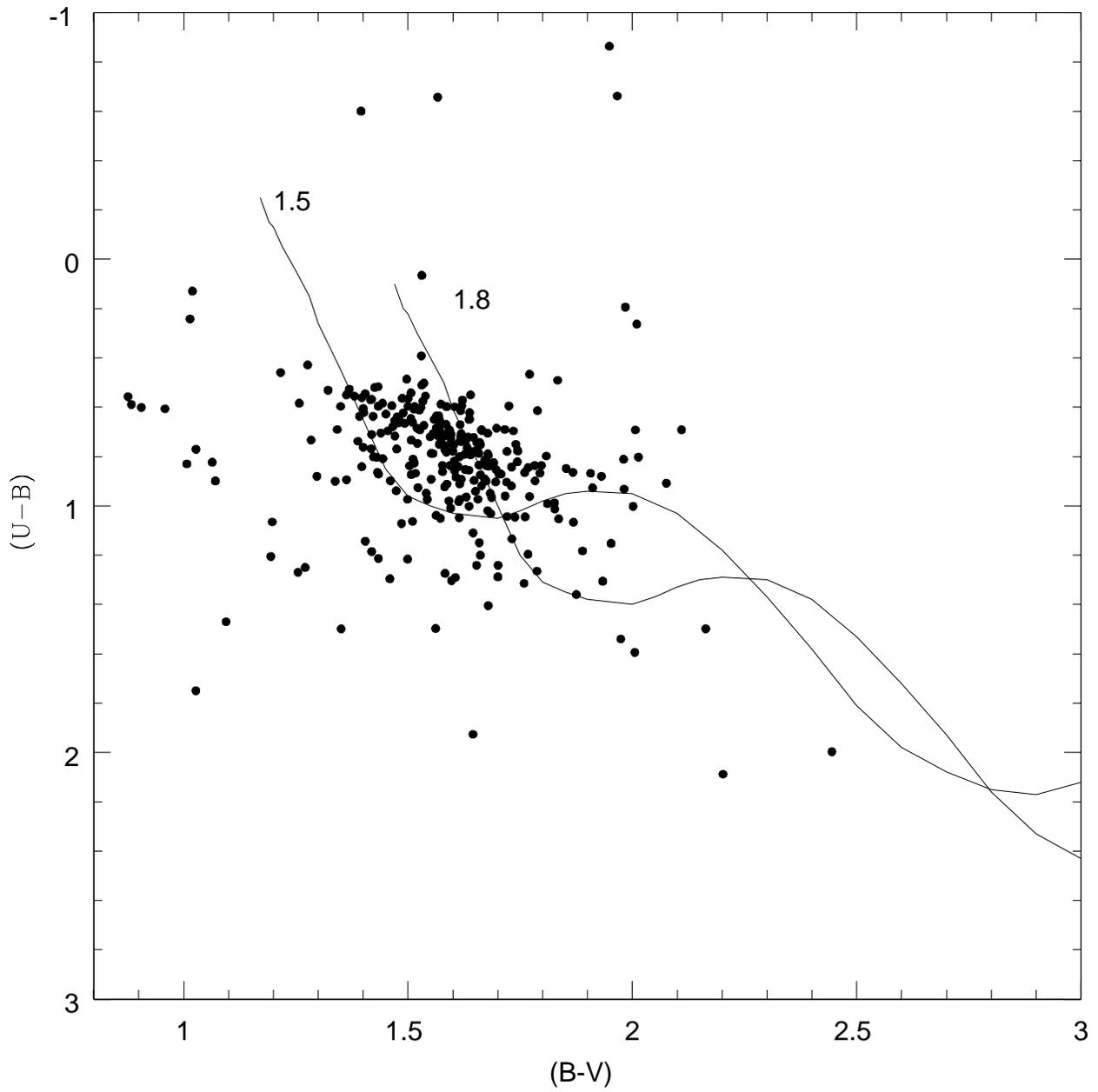}
\caption{Estimation of reddening towards the cluster using all the stars observed in the 
cluster region using (U$-$B) vs (B$-$V) colour- colour diagram.}
\end{figure*}

%
%
\begin{figure*}
\epsfxsize=18truecm
\epsffile{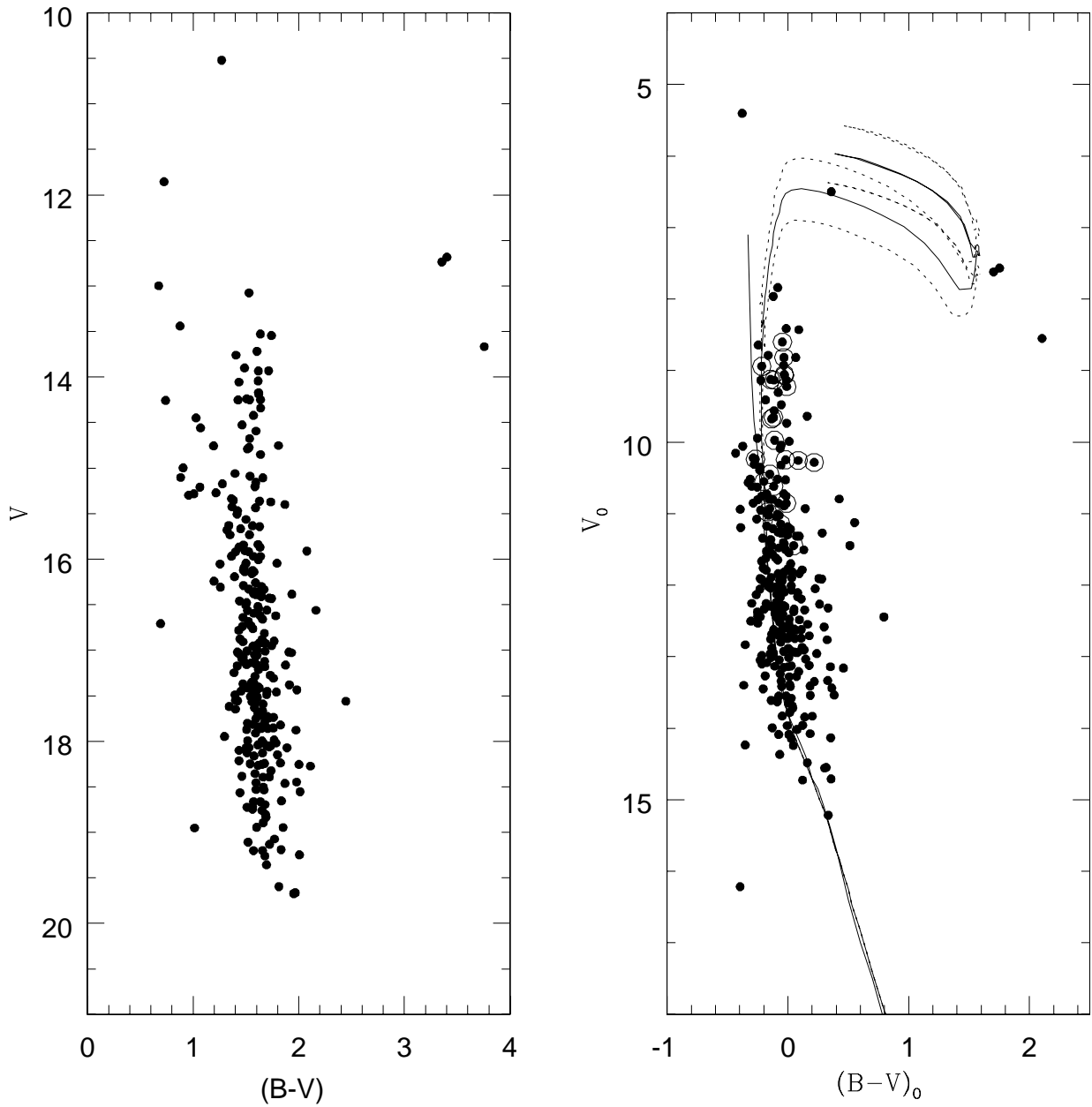}
\caption{The observed CMD of 327 stars is shown in the left panel. The reddening and extinction corrected CMD  
after removing the foreground stars is shown in the right panel. Fit of ZAMS and isochrone
to the cluster MS and turn-off is also shown. The estimated distance modulus is 12.3 mag. The age of the
isochrones shown are 25 Myr (bold line), 20 and 32 Myr (dashed lines).  The emission line stars are denoted 
by open circles around dots.}
\end{figure*}

%
\begin{figure*}
\epsfxsize=18truecm
\epsffile{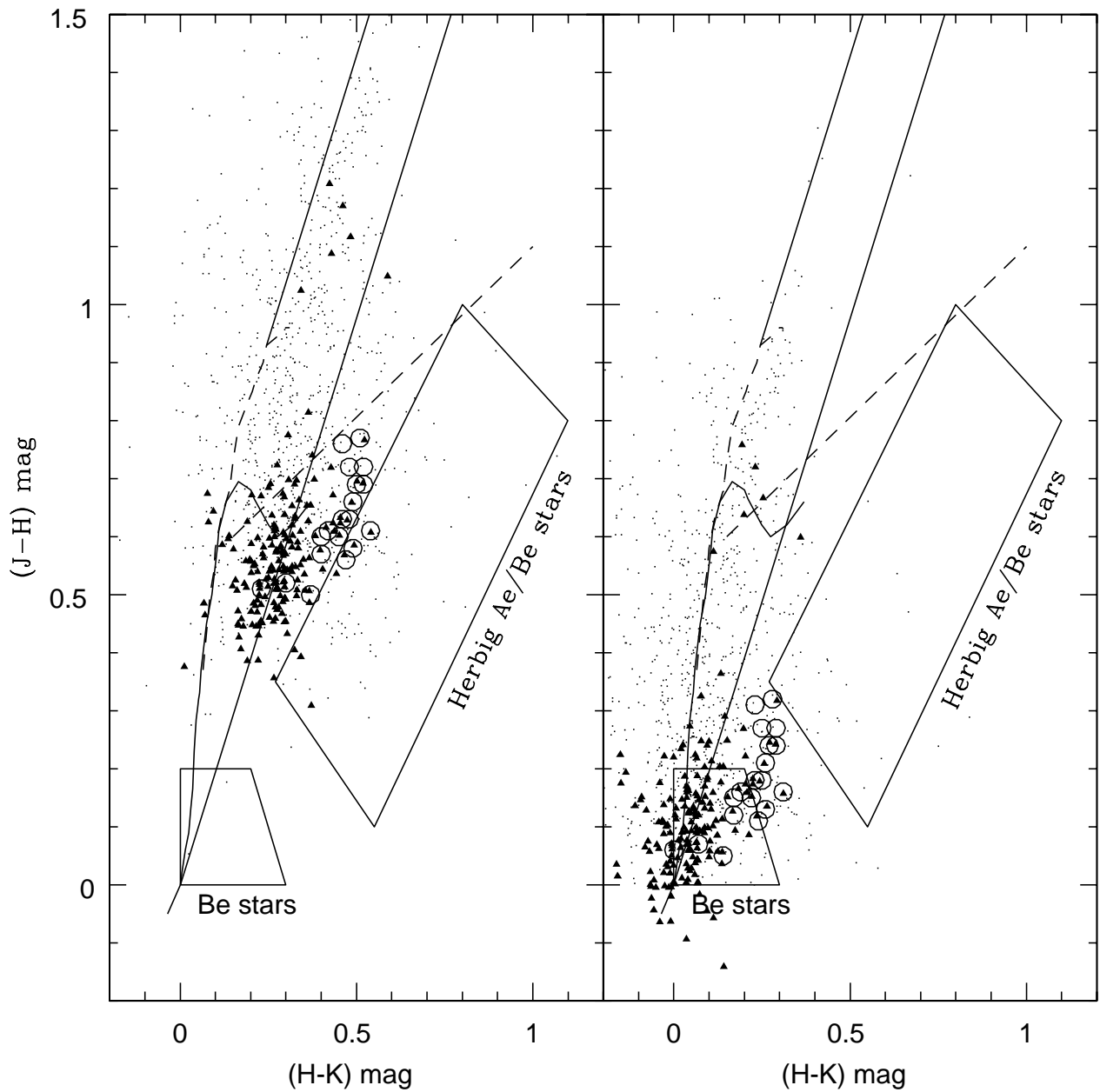}
\caption{The NIR colour-colour diagram for stars in the region of the cluster is shown in the left panel. 
The small dots
indicate all detections in the observed region, triangles indicate stars with optical counterparts, 
large open circles indicate emission line stars. The location
of Herbig Ae/Be stars and classical Be stars are also indicated. The location of stars in the diagram
after reddening correction is shown in the right panel.}
\end{figure*}

%
%
\begin{figure*}
\epsfxsize=18truecm
\epsffile{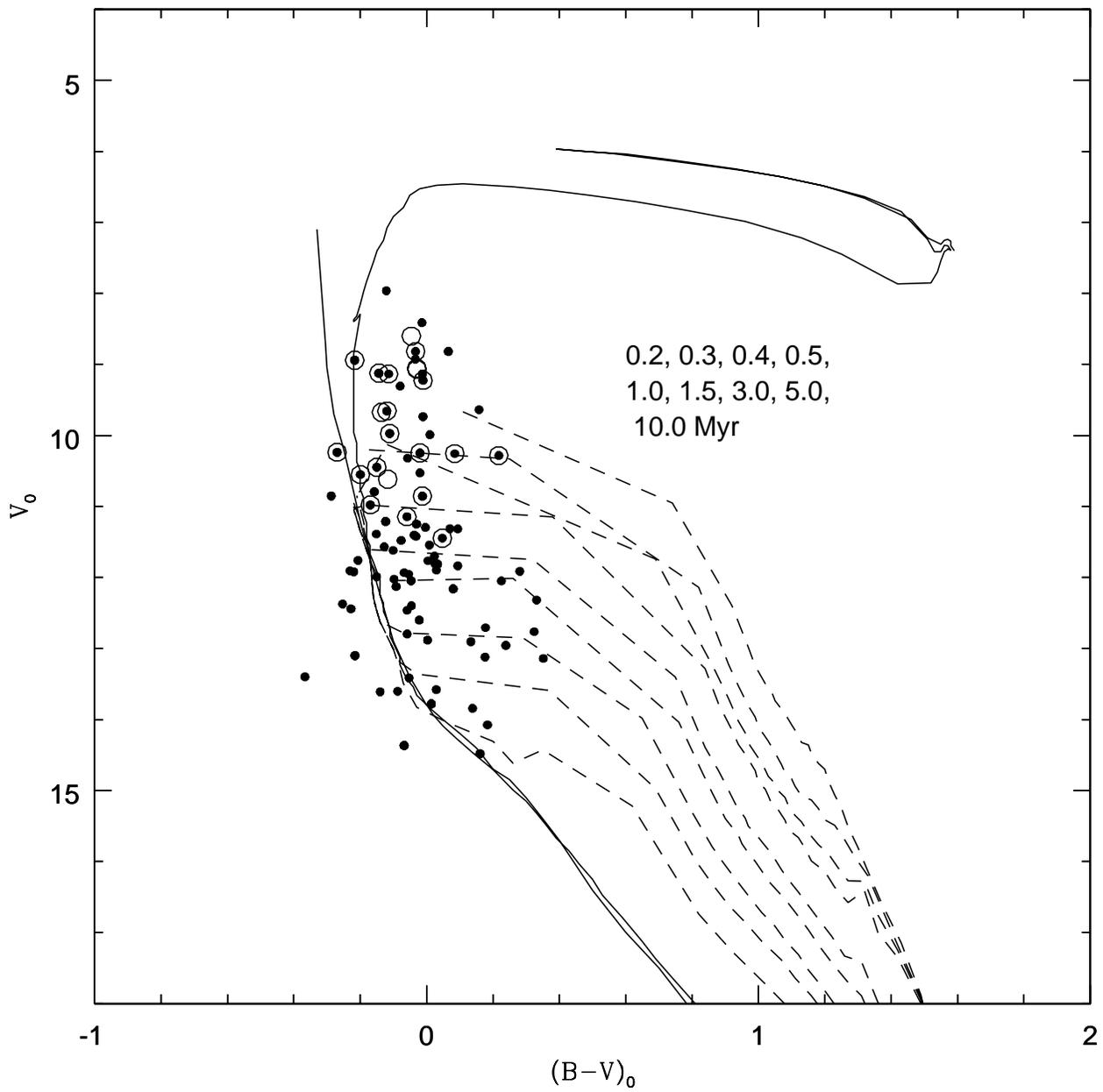}
\caption{The optical CMD of stars with NIR excess. The dots with circles around them are the emission line
stars. The pre-MS isochrones of Siess et al. (2000) for various ages are shown in dashed lines.
The post-MS isochrone of age 25 Myr and ZAMS are shown in continuous lines.}
\end{figure*}

%
%
\begin{figure*}
\epsfxsize=18truecm
\epsffile{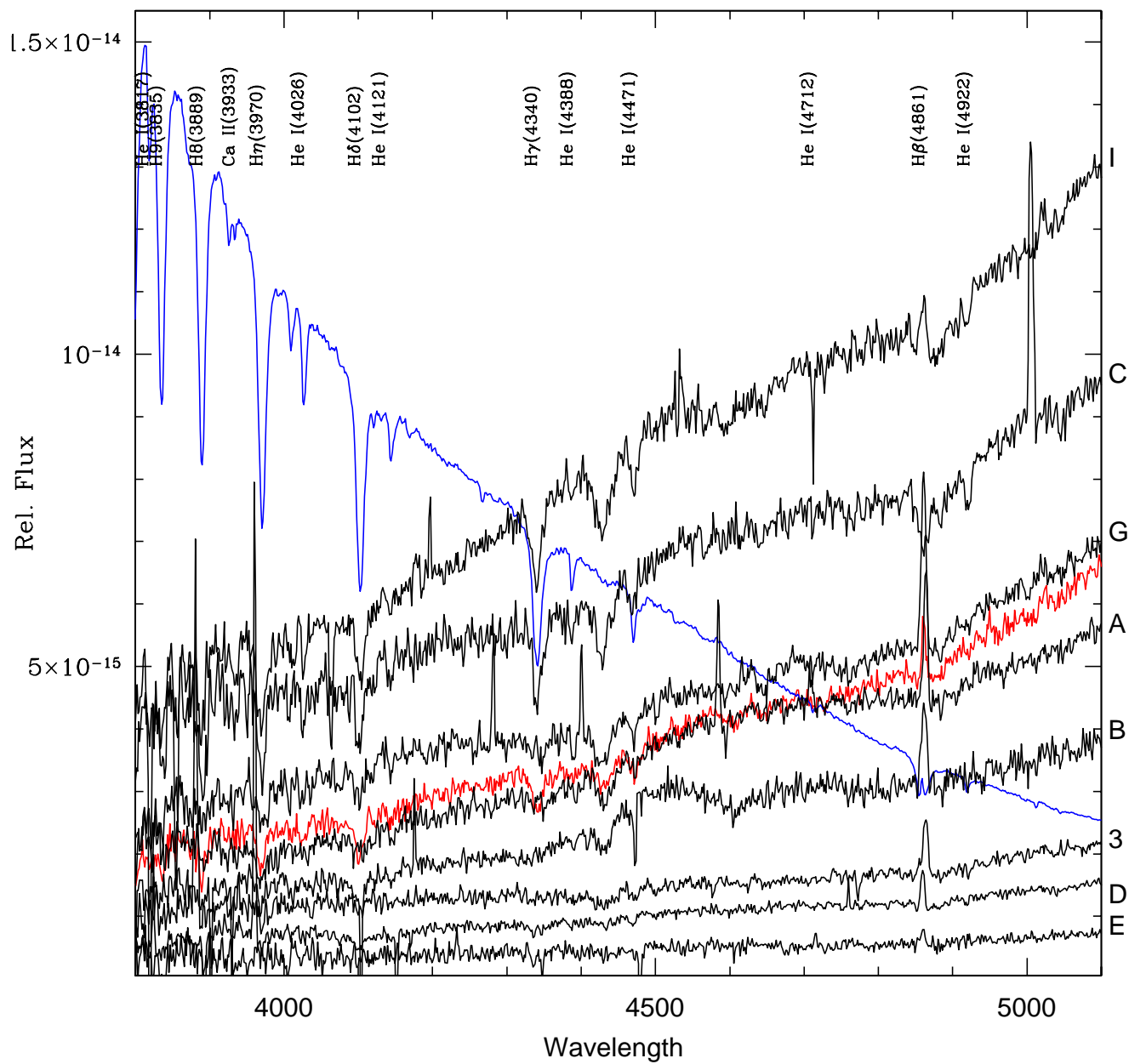}
\caption{Low resolution spectra of stars A,B,C,D,E,G,I and 3. Prominent spectral lines are
also shown. For comparison, the spectra of HBC 705, a B2 Herbig Be star (red) and HD 58343, a B3Ve, classical
emission line star (blue) are also plotted.}
\end{figure*}

%
%
\begin{figure*}
\epsfxsize=18truecm
\epsffile{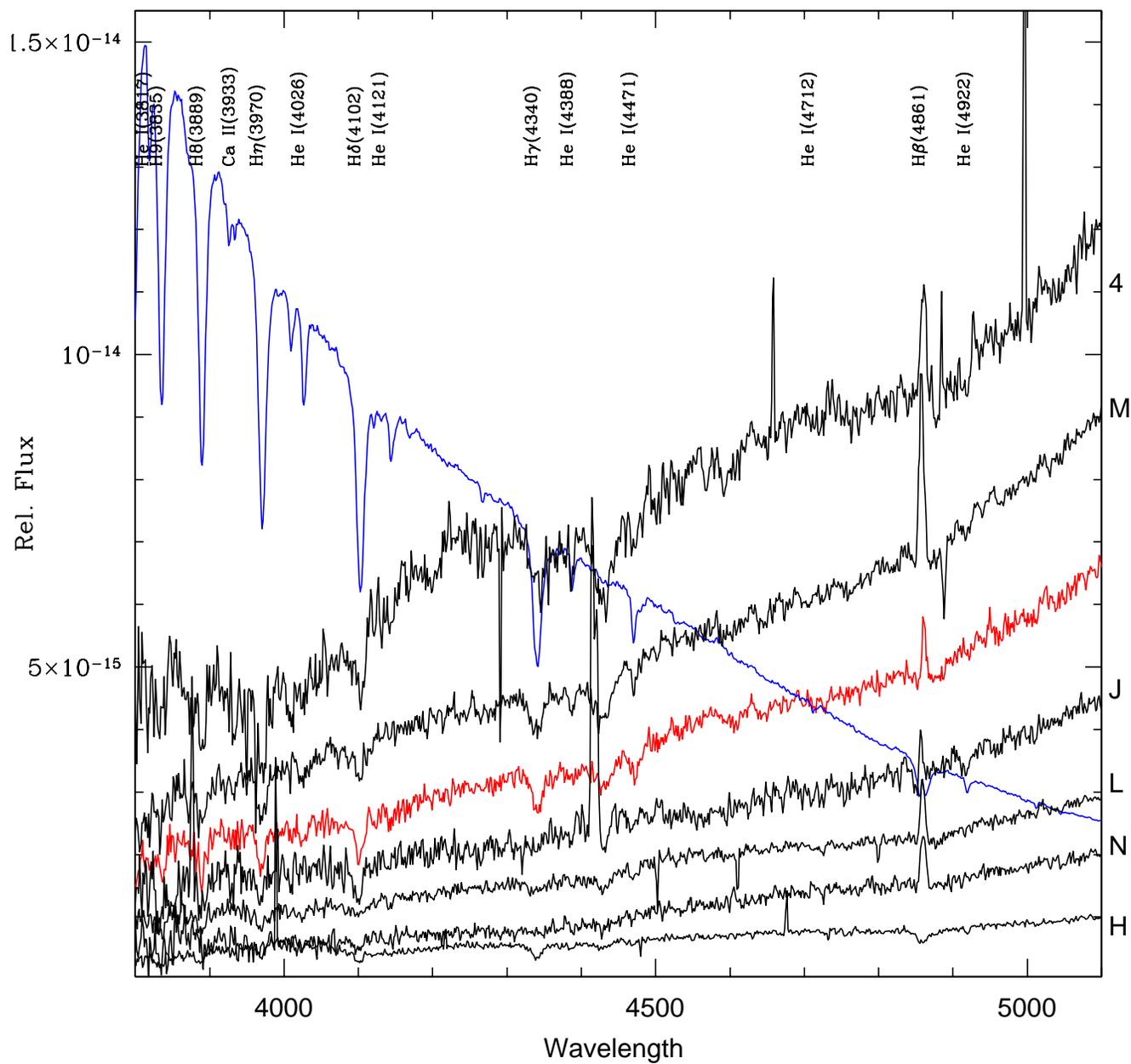}
\caption{Same as figure 6, but for stars, H,J,L,N,M and 4.}
\end{figure*}

%
%
\begin{figure*}
\epsfxsize=18truecm
\epsffile{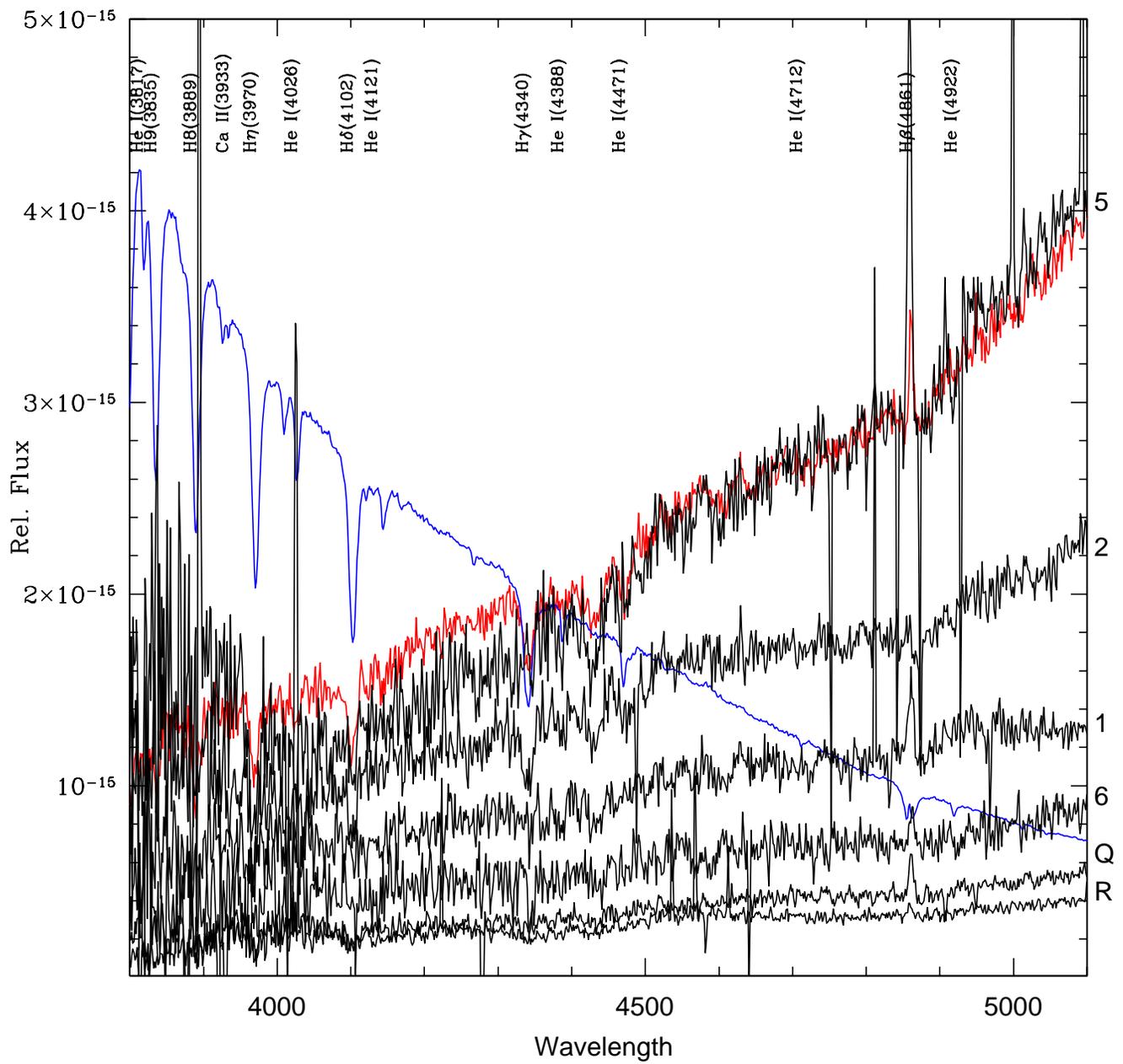}
\caption{Same as figure 6, but for stars Q,R,1,2,5 and 6. }
\end{figure*}
%
%
\begin{figure*}
\epsfxsize=18truecm
\epsffile{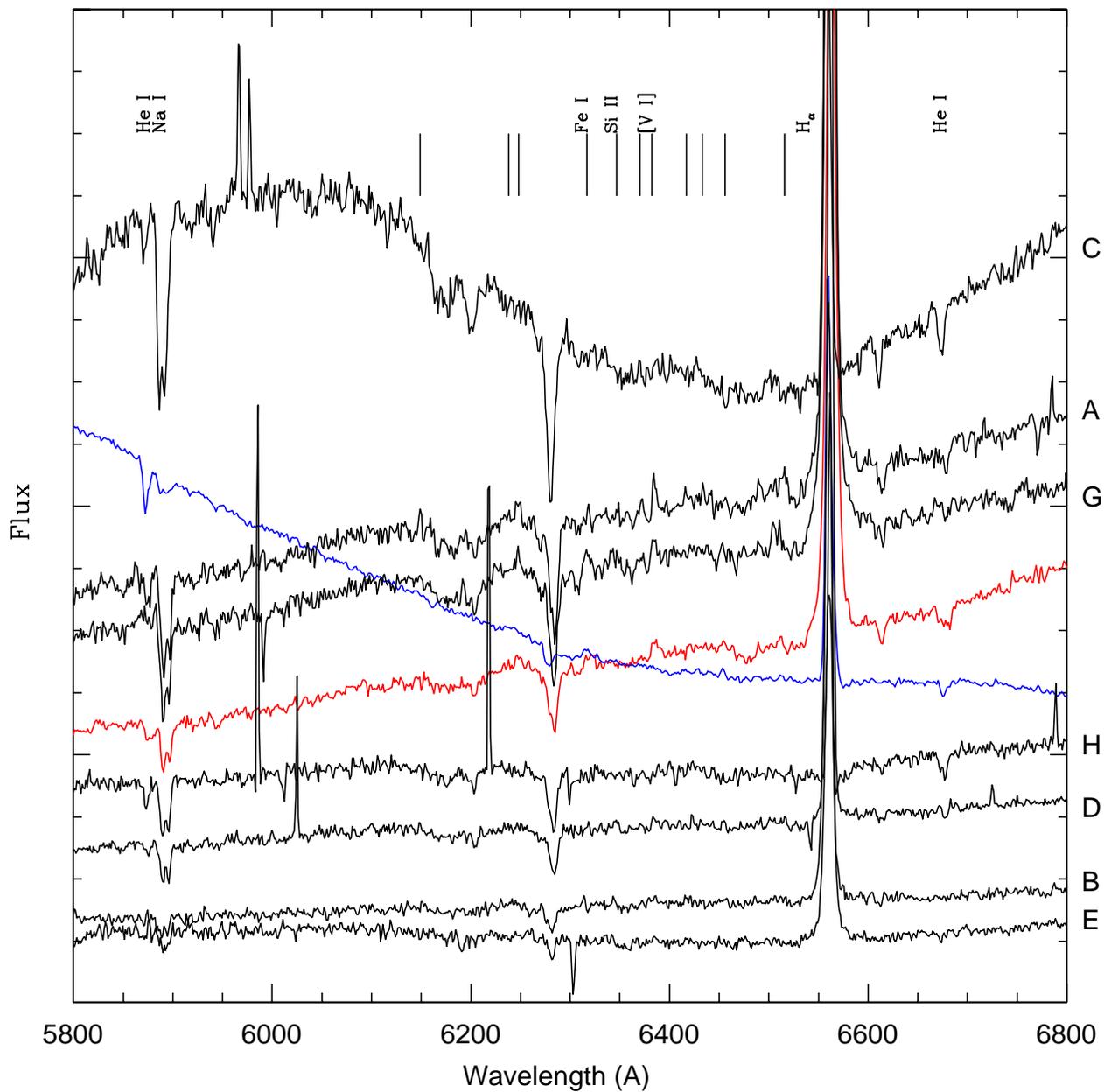}
\caption{Low resolution spectra in H$_\alpha$ region. The spectra are labelled. The red and the blue
spectra indicate HBC 705 and HD 58343 respectively.}
\end{figure*}
%
%
%
%
%
%
%
\begin{figure*}
\epsfxsize=18truecm
\epsffile{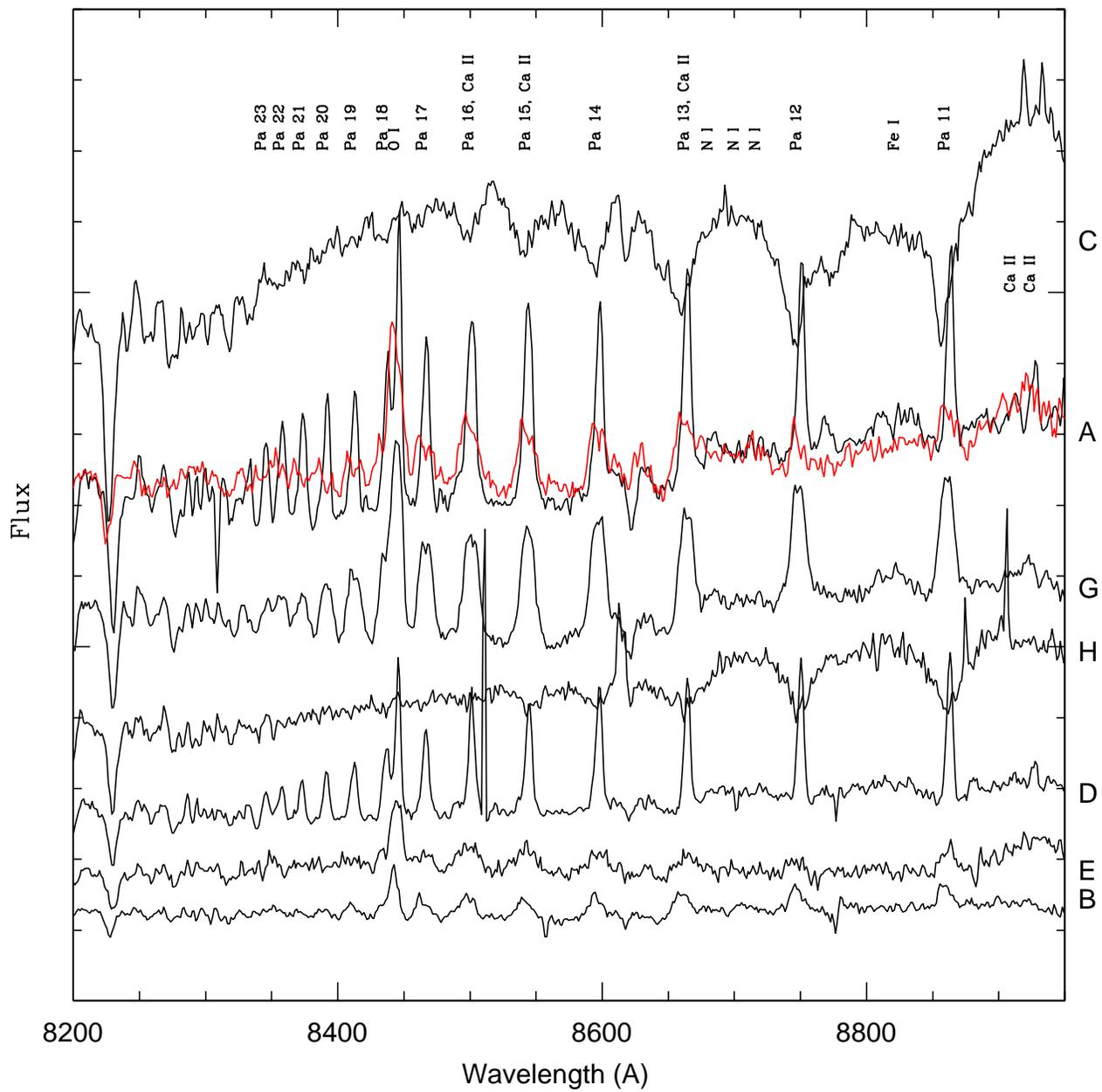}
\caption{Low resolution spectra of emission line stars near the red region. The spectra are labelled and
prominent lines are listed. The spectrum of HBC 705 is shown in red, for comparison. }
\end{figure*}
%
%
\begin{figure*}
\epsfxsize=18truecm
\epsffile{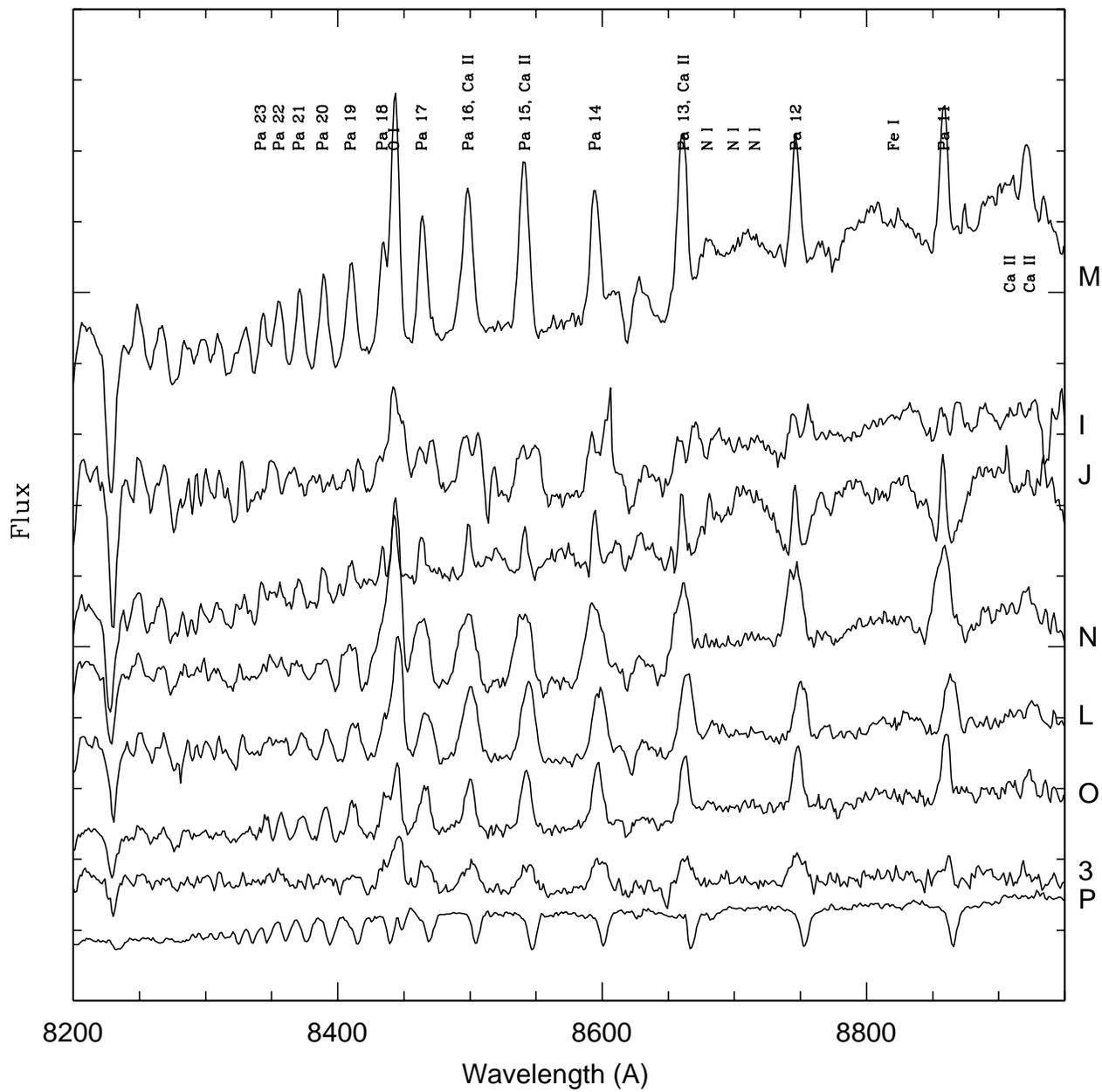}
\caption{Same as figure 13,without the spectrum of HBC 705.}
\end{figure*}
%
%
\end{document}